\documentclass[a4paper,twocolumn]{revtex4-1}
\usepackage[latin1]{inputenc}
\usepackage{graphicx}
\usepackage{verbatim}
\usepackage{amsmath}
\usepackage{amsfonts}
\usepackage{amssymb}
\usepackage{latexsym}
\newcommand{\beq}{\begin{equation}}
\newcommand{\eeq}{\end{equation}}

\begin{document}
\title{A simple model for the distribution of plasmid lengths }
\author{Alice Ledda}
\affiliation{Infection, Antimicrobiens, Mod{\'e}lisation, Evolution (IAME), 
UMR 1137 INSERM, Universit{\'e}s Paris Diderot et Paris Nord, 
Sorbonne Paris Cit{\'e}, Facult{\'e} de M{\'e}decine, 
Site Xavier Bichat, 16 rue Henri Huchard 75018, Paris, France}
\email{alice.ledda@gmail.com}

\author{Luca Ferretti}
\affiliation{Syst\'ematique, Adaptation et Evolution (UMR 7138), UPMC Univ Paris 06, CNRS, MNHN, IRD, Paris, France}
\affiliation{CIRB, Coll\`ege de France, Paris, France}
\begin{abstract}
Plasmids are major players in Horizontal Gene Transfer mechanisms, hence they are highly variable in their gene content and length. We propose a model for the fitness of a plasmid as a function of its length, which predicts diminishing returns. We infer the distribution of plasmid lengths by a simple evolutionary model and we show that there is a positive correlation between the presence and efficiency of the conjugation machinery and the length of the plasmid. The model predicts an ``insertion load'' on plasmids, which could explain also the fact that plasmids are widespread in bacterial populations but rarely established. Finally we discuss how the typical length of plasmids increases with the amount of stress in the environment, focusing on the recent human-driven increase in antibiotic concentrations. 
\end{abstract}
\maketitle

\section*{Introduction}

Plasmids are DNA elements ubiquitous in all types of bacteria and archaea. They are mainly constituted of non-housekeeping genes that perform very specific functions like virulence factors, antibiotic and heavy metal resistances, apart form genes encoding for selfish functions explicitly aimed at plasmid survival and/or dissemination like active partitioning system, addictive complexes and replication/transmission machinery \cite{funnell2004plasmid,norman2009conjugative}. Often plasmid encoded genes are not necessary for cell survival in average conditions (i.e. no stress, no starvation). Yet sometimes they are determinant for cell survival in specific environmental conditions that can  appear suddenly, i.e. when antibiotic is introduced in the environment \cite{svara2011evolution}

Most plasmids can be inherited both vertically from parent cells and horizontally from fellow bacteria (not necessarily belonging to the same species or genus), through a mechanism called conjugation \cite{funnell2004plasmid}. 
 They are classified by their mobility abilities as conjugative, mobilizable and non-transmissible \cite{smillie2010mobility}. Conjugative plasmids encode all the machinery to be transmitted into another bacteria, which is made from a relaxase and a Mating Pair Formation (MPF) complex that allows mating pair formation (physically attaching one bacteria to another) and hence plasmid transmission from one bacteria to another. Mobilizable plasmids encode the relaxase but need to rely on mating pair machinery from another plasmid, while non-transmissible plasmids do not encode any gene useful for plasmid transmission \cite{smillie2010mobility, carattoli2009resistance, francia2004classification, garcillan2011identification}.  
 Plasmid mobility is a key factor to understand the population genetics of plasmid encoded genes such as antibiotic resistances or virulence factor \cite{harrison2012plasmid, nogueira2009horizontal, davies2010origins, carattoli2009resistance}.
 
 Many population genetics studies investigated the conditions for plasmids survival in bacterial populations in environments where they do not enhance bacterial fitness. The key factor is the trade-off between plasmids conjugation rates and plasmid cost for bacteria \cite{stewart1977population, hoeven1984mathematical, dionisio2005evolution, tazzyman2013fixation, levin1977probability,bichsel2010early,bichsel2013estimating}. 

Due  to their ability of conjugating between bacteria, plasmids are vectors of horizontal gene transfer.  Yet they are themselves subject to horizontal gene transfer as their gene content is shaped from Insertion Sequences (IS) from host chromosomes \cite{carattoli2009resistance, norman2009conjugative}. We can think of conjugative plasmids as a primitive sequence constituted from the conjugation machinery that grows with the insertion of ISs. The rate at which new IS are acquired is not clear, but is assumed to be low so that the plasmid can spread around in different populations and environments before a new IS is added. It is clear that even if the gene pool from which the IS can acquire genes is ideally infinite, it can be summarized by a finite number of functions performed from these genes. 

We propose 
that plasmids evolve to include genes to span as many different functions as possible: the more functions they perform, the more probable it is that they increase the fitness of the host bacteria by performing a life-saving function and therefore they will not be lost. On the other hand, each gene encoded on the plasmid represents a fixed cost for bacteria, so redundancy in gene functions will be avoided as much as possible, as having many genes performing the same function will result in an higher cost but no higher benefit compared to having a single gene performing the same function. 

The aim of this article is to study how the number of genes and the functions they perform shape plasmid length. We model the average fitness of plasmids of a given length and we include it in a simple evolutionary model to infer the distribution of plasmid lengths. We study the relation between the properties of the conjugation machinery and the typical length of the plasmid. Finally, we discuss the effect of the recent increase of antibiotic concentration in the environment driven by human use\cite{andersson2014microbiological}

\section*{Methods}

\subsubsection*{A model of length-dependent plasmid fitness}

Plasmids are transmitted both  horizontally (i.e.  by conjugation) or vertically (by duplication of the host bacterium), therefore the plasmid fitness is a combination of the plasmids contribution to the host fitness  and its own transmission rate.

The pan-genome of plasmids contains many genes with several, possibly overlapping, functions. Insertion of new genes can have a reduced fitness advantage in long plasmids, since these genes could share some functions with the preexisting genes and therefore be partly or totally redundant. This phenomenon can be described by the classical ``coupon collector'' model \cite{coupon}. Functions of new genes are extracted at random from a predefined set of $n^*_f$ functions, where each function gives a fitness advantage $s^*/n^*_f$ to the plasmid. The number of functions $n_f$ present in the plasmid grows on average according to the equation
\beq
\frac{\Delta n}{\Delta l}=\frac{1}{l_{gene}} \left(1-\frac{n_f}{n^*_f}\right)
\eeq
where $1-{n_f}/{n^*_f} $ is the probability that the function is already present in the plasmid sequence. Since most plasmids are at least tens of genes long [cite], we can solve the above equation in the continuous approximation. Defining $k=1/n^*_fl_{gene}$, the average fitness advantage for $l/l_{gene}$ genes is given by $\Delta f=\frac{s^*}{n^*_f}n_f$, i.e. 
\beq
\Delta f(l)=s^*(1-e^{-kl})
\eeq
where $s^*$ is the maximum fitness advantage and $k$ is inversely proportional to the number of different functions.

On the  other hand, since the bacteria have to replicate and transcribe/translate plasmid genes, the fitness cost for new genes is proportional to the length: $\Delta f(l)=-cl$, where $c$ is the cost per unit length. 

Finally, many plasmids are able to transfer a 
copy of themselves into other bacteria, a process called conjugation, which plays an important role in Horizontal Gene Transfer (HGT) across bacteria. It has been shown that this process gives an effective additive fitness contribution to the plasmid equal to the conjugation rate $r_c$ \cite{stewart1977population, hoeven1984mathematical, dionisio2005evolution, tazzyman2013fixation, levin1977probability,bichsel2010early,bichsel2013estimating}. Conjugative plasmids contain genes responsible for the complex conjugation machinery, which often come from a single HGT event from a bacterial genome. These genes are transferred or inherited together, therefore it makes sense to consider them as a single block of length $l_c$ and fitness cost $c_c$ (including the cost of the machinery). 

Therefore, summing all the contributions above, the fitness of plasmids of length $l$ has the form:
\beq
f(l)=s^*(1-e^{-k(l-l_c)})-c(l-l_c)+r_c-c_c\label{eq_fitness0}
\eeq
and we can rewrite it in terms of four parameters $s=s^*e^{kl_c}$, $k$, $c$ and $r=r_c-c_c+s^*(e^{kl_c}-1)+cl_c$, obtaining:
\beq
f(l)=s(1-e^{-kl})-cl+r\label{eq_fitness}
\eeq


Note that even if we add extra sets of genes with a fixed cost $\tilde{c}$, fitness advantage $\tilde{s}$ and length $\tilde{l}$, the shape of the fitness curve remains the same up to 
a redefinition of the parameters:
\beq
\begin{split}
f(l)=\tilde{s}-\tilde{c}+s(1-e^{-k(l-\tilde{l})})-c(l-\tilde{l})+r=\\ [se^{k\tilde{l}}]\cdot(1-e^{-kl})-cl+[r+\tilde{s}-\tilde{c}+c\tilde{l}-s(e^{k\tilde{l}}-1)]
\end{split}
\eeq

\subsubsection*{A model of plasmid evolution}

We consider a model of plasmid evolution that includes two processes: reproduction/transmission and insertion of new genes. New plasmids are born at constant rate at a minimum length $l_0>l_c$. Existing plasmids can reproduce at a rate given by their fitness $f(l)$ or increase their length by insertion of new genes at rate $\mu(l)$.  

We ignore the stochastic fluctuations in both processes. Fluctuations in the reproduction rate can be neglected for large population sizes, while exact equations which include the fluctuations in the insertion rate will be considered in the Appendix. The approximate deterministic equation for the evolution of the number of plasmids of a given length is
\beq
\frac{\partial n(l,t)}{\partial t}=f(l)n(l,t)-\frac{\partial }{\partial l}[\mu(l)n(l,t)]\label{eq:diffeq}
\eeq
and the stationary solution can be found either by the method of characteristics or by direct integration of ${\partial n(l)}/{\partial t}=0$:
\beq
n(l)\propto \frac{1}{\mu(l)}\exp\left(\int_{l_0}^l dx \frac{f(x)}{\mu(x)}\right)\label{eq:nl}
\eeq

In our model, we assume that the insertion rate $\mu$ does not depend on the length of the plasmid and that the fitness $f(l)$ is the one obtained in equation (\ref{eq_fitness}). 

If new plasmids are born with a distribution of initial lengths $p_0(l_0)$, the solution becomes 
\beq
n(l)\propto \frac{1}{\mu(l)}\int dl_0 p_0(l_0)\exp\left(\int_{l_0}^l dx \frac{f(x)}{\mu(x)}\right)\label{eq_fitness_p0}
\eeq

Finally, the stationary distribution of plasmids length is given by $p(l)=n(l)/\int_{l_0}^\infty n(x) dx$.

If the distribution is sharply peaked, it can be approximated around the peak using a second order Taylor expansion of the exponent in equation (\ref{eq:nl}). The result is a Gaussian distribution $p(l) \propto \exp\left(-\left|\frac{\partial }{\partial l}\left(\frac{f(l)}{\mu(l)}-\frac{1}{\mu(l)}\frac{\partial \mu(l)}{\partial l}\right)\right|_{l=l_{peak}} (l-l_{peak})^2\right)$ with mean $l_{peak}$ and variance 
\beq
\mathrm{Var}(l)\simeq-\frac{1}{2} \left[ \left.\frac{\partial}{\partial l} \left(\frac{f(l)}{\mu(l)}-\frac{1}{\mu(l)}\frac{\partial \mu(l)}{\partial l}\right)
\right|_{l=l_{peak}} \right]^{-1}
\eeq

\section*{Results}

\subsubsection*{Fitness as a function of plasmids length}
\begin{figure}
\begin{center}
\includegraphics[width=\columnwidth]{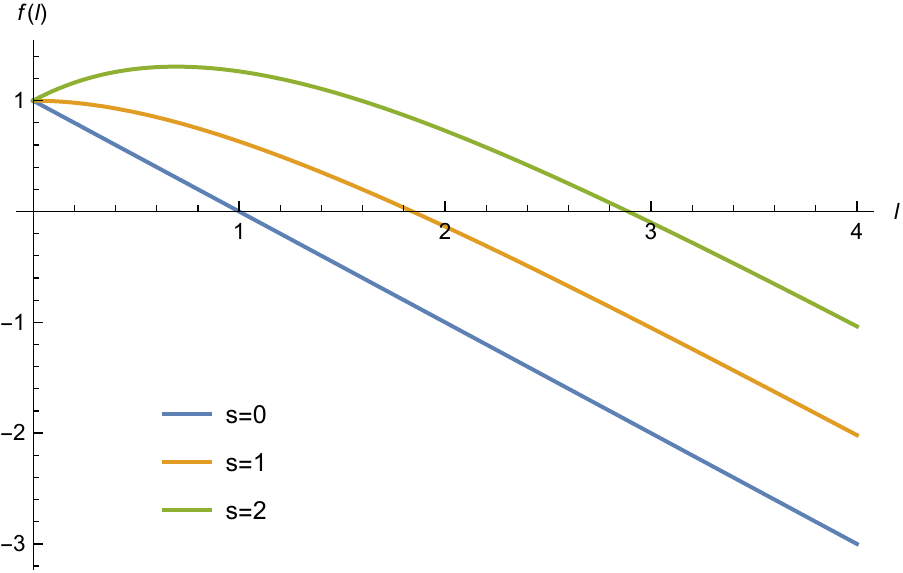}
\caption{Fitness function $f(l)$, as a function of the length $l$, for
different values of the selection advantage $s$. The other parameters
are $c=k=r=1$.}\label{fig_fitness}
\end{center}
\end{figure}

\begin{figure}
\begin{center}
\includegraphics[width=\columnwidth]{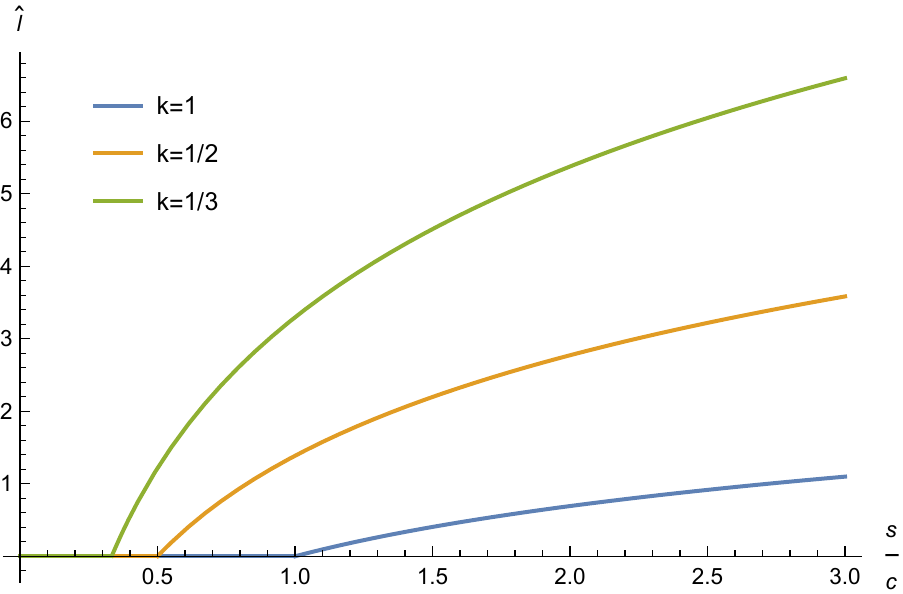}
\caption{Location of the length $\hat{l}$ of maximum fitness, as a
function of the selection/cost ratio $s/c$, for different values of
$k$. The location of the maximum does not depend on the parameter
$r$.}\label{fig_fmax}
\end{center}
\end{figure}

The plasmid fitness function is given in equation (\ref{eq_fitness}) and shown in Figure \ref{fig_fitness}. It can have two different behaviours depending on the parameter $\rho=kse^{-kl_0}/c$, which represents the ratio between benefits and costs of the first gene added to the plasmid. If $\rho<1$, the fitness is a decreasing function of plasmid length and it is maximal at the initial length $\hat{l}=l_0$ (blue curve in Figure \ref{fig_fitness}). If $\rho>1$, the fitness increases with plasmid length until a maximum at 
\beq
\hat{l}=\frac{1}{k}\log\left(\frac{ks}{c}\right) 
\eeq
and then drops down (green curve in Figure \ref{fig_fitness}). 

The location of the maximum depends on the parameters and its behaviour is shown in Figure \ref{fig_fmax} as a value of the ratio between gene benefit and cost $s/c$ for different values of $k$. Note that the maximum is not dependent on the plasmid conjugation rate $r$. 
The length of maximum fitness is zero for  $s/c< k/e^{-kl_0} $ and then grows with $s/c$, while the slope of the growth decreases with $k$.

This fitness function is concave for all choices of parameters, 
so the insertion of further genes tends to have a ``diminishing return'', i.e. a smaller effect on the total fitness of the plasmid.

\subsubsection*{The distribution of plasmid lengths}

\begin{figure}
\begin{center}
\includegraphics[width=\columnwidth]{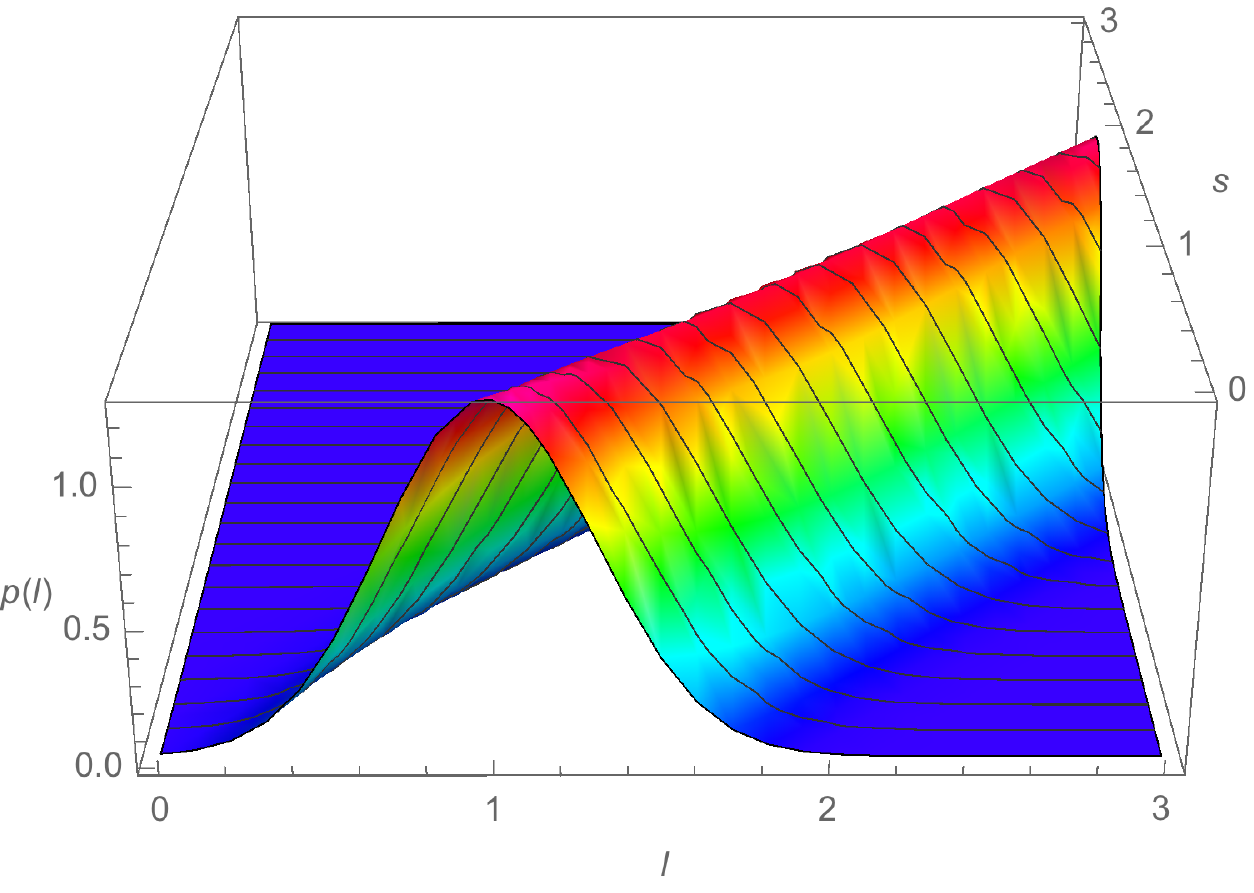}
\caption{Length distribution $p(l)$ as a function of length $l$ and
selection advantage $s$, for different values of the selection
advantage $s$. The other parameters are $c=k=r=1$,
$\mu=0.01$, $l_0=0$.}\label{fig_rift}
\end{center}
\end{figure}

\begin{figure}
\begin{center}
\includegraphics[width=\columnwidth]{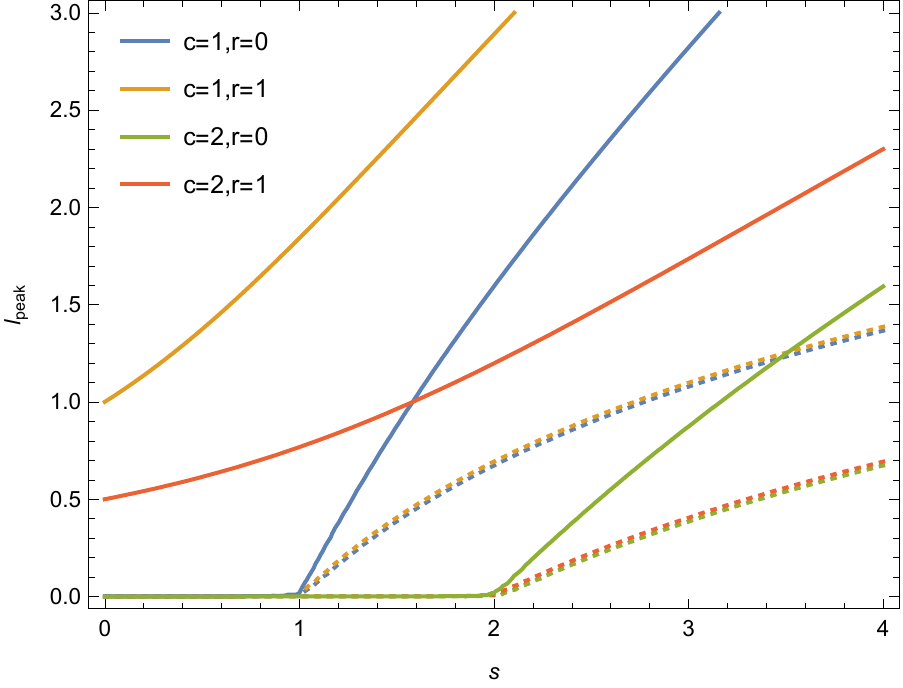}
\caption{Most probable length $l_{peak}$ of a plasmid, as a function
of the selection advantage $s$, for different values of $c$ and $r$.
For comparison, the dashed lines show the location of the fitness
maximum for the same parameters. Other parameters are $k=1$, $l_0=0$. }\label{fig_dmax}
\end{center}
\end{figure}

Our prediction for the plasmid length distribution is
\begin{widetext}
\beq
p(l)\propto\exp \left[\frac{1}{\mu}\left( (r+s)(l-{l_0})-\frac{ s}{k}
\left(e^{-k l_0}-e^{-k l}\right)-
\frac{c}{2}(l^2-{l_0}^2)\right)\right]\label{eq:p}
\eeq
\end{widetext}
Its behaviour with $s$ in shown in Figure \ref{fig_rift}.

The peak of the distribution $l_{peak}$ corresponds to the most
probable plasmid length. The condition for the maximum $p'(l_{peak})=0$
turns out to be 
\beq
f(l_{peak})=0 \label{eq:f0}
\eeq
hence the most probable
plasmid length $l_{peak}$ is one of the zeros of the fitness function. This means that typical plasmids tend to reach equilibrium in bacterial populations, instead of invading or disappearing.

For the fitness (\ref{eq_fitness}), the analytical solution in terms of the Lambert $W$ function (the inverse function of $xe^x$) is
\beq
l_{peak}=\max\left[l_0,\frac{s+r}{c}+\frac{1}{k}W\left(-\frac{ks}{c} e^{-\frac{k(s+r)}{c}}\right)\right]
\eeq
The dependence of $l_{peak}$ on the parameters is illustrated in
Figure \ref{fig_dmax}. It is striking to note that the average plasmid length is consistently much 
larger than the length that maximises the plasmid fitness. This result $l_{peak}\geq \hat{l}$ can be retrieved analytically and it is grounded on the properties of equation (\ref{eq:p}). The difference $\Delta_{load}=l_{peak}-\hat{l}$ represent the length of the amount of genes in  excess with respect to the optimal fitness. These genes give a negative fitness load 
\beq
f_{load} =\begin{cases}  -s-r+\frac{c}{k}(1+\log\left(\frac{sk}{c}\right)) &\mathrm{for}\ \rho>0 \\
-s(1-e^{-kl_0})+cl_0-r & \mathrm{for}\ \rho<0
\end{cases}
\eeq

\begin{figure}
\begin{center}
\includegraphics[width=\columnwidth]{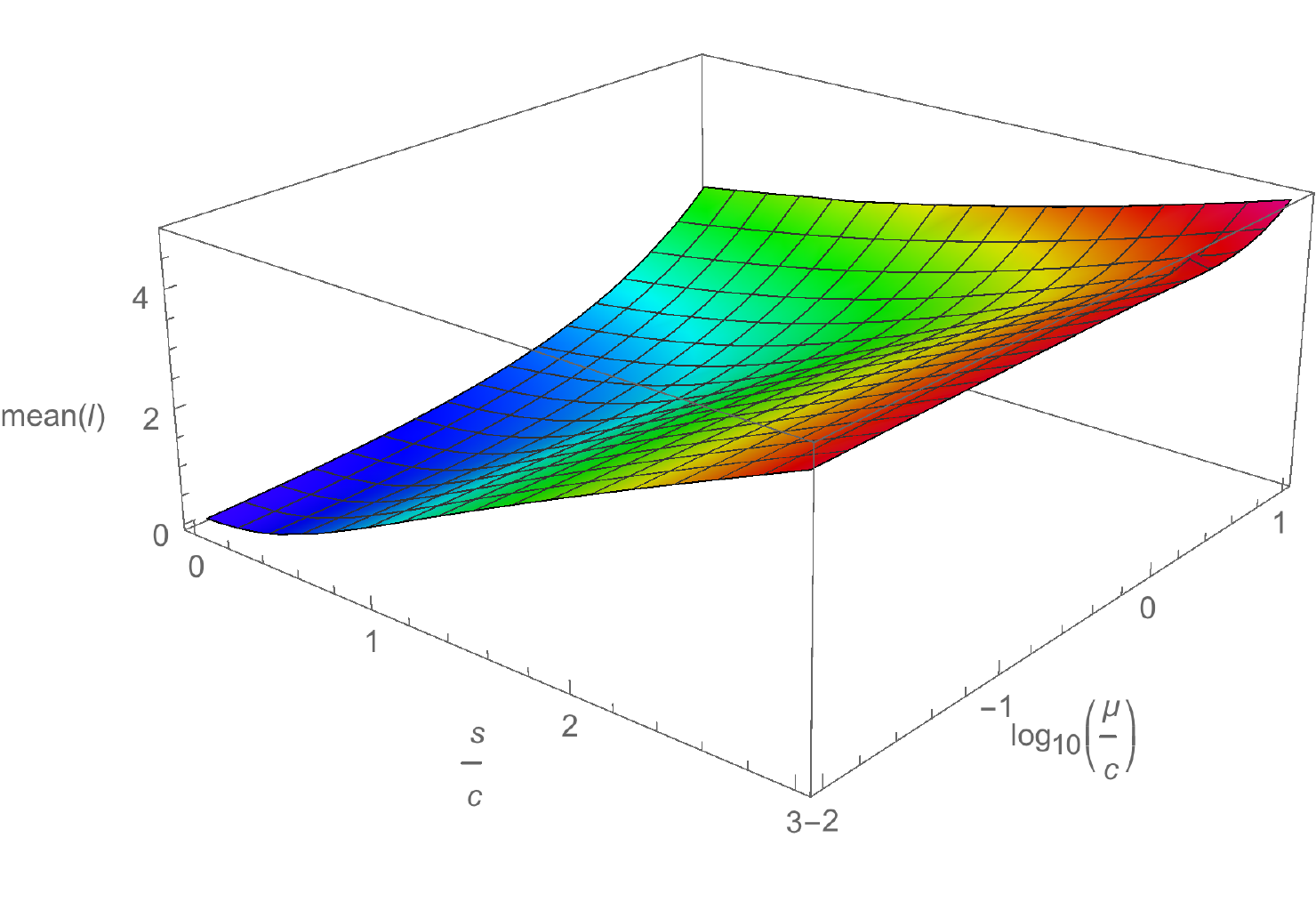}
\includegraphics[width=\columnwidth]{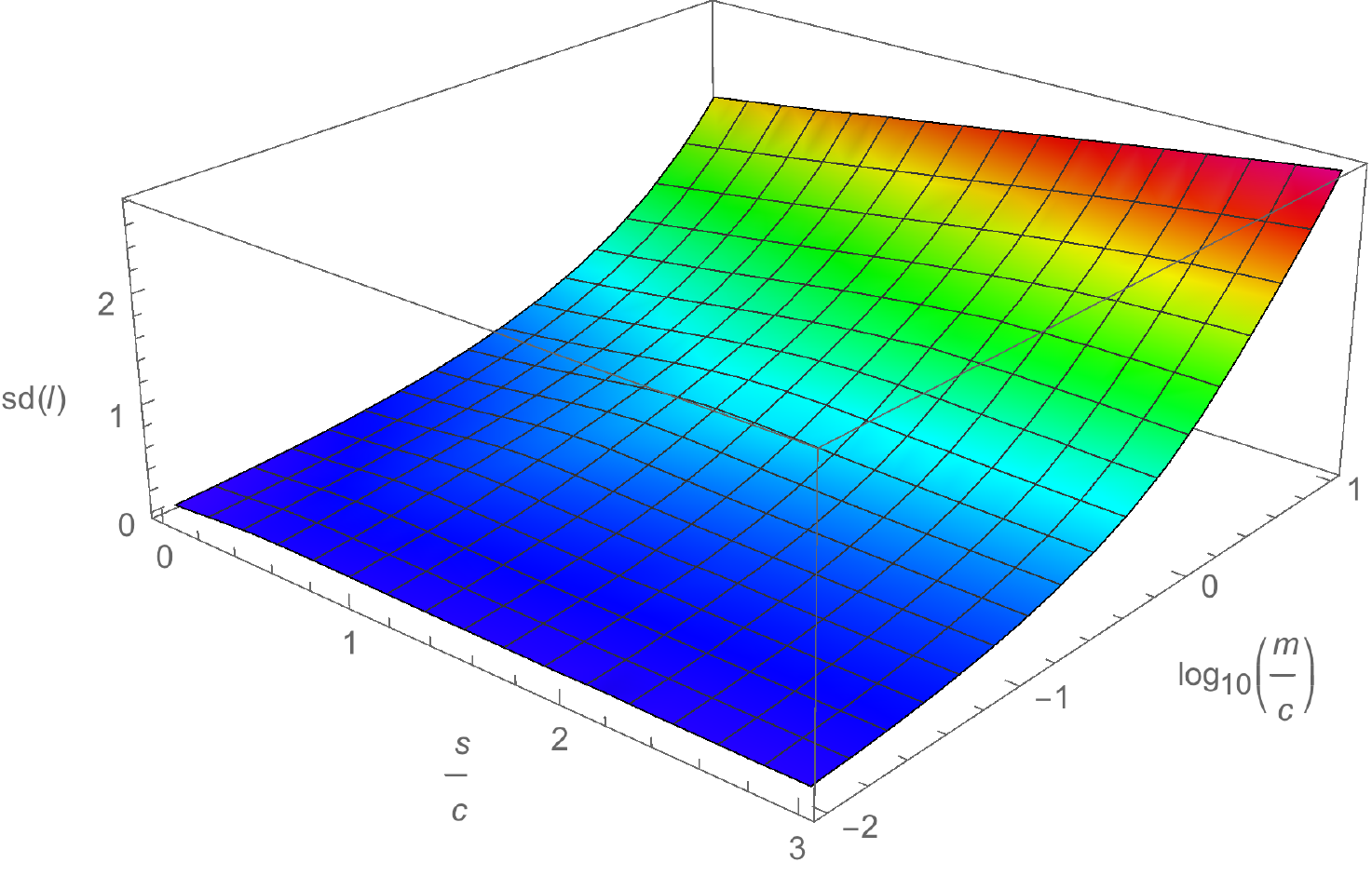}
\caption{Mean and standard deviation of the distribution of plasmid
lengths, as a function of the benefit/cost ratio $s/c$ and the
insertion/cost ratio $\mu/c$.  The other parameters are $r=s/2$,
$k=1$, $l_0=0$.}\label{fig_sd}
\end{center}
\end{figure}

The behaviour of the length fluctuations $\sigma(l)$ is illustrated in Figure \ref{fig_sd}. The width of the length distribution grows with the insertion
rate $\mu$ and gets smaller for larger benefits $r$, $s$. In fact, for $\rho>0$, the width of the length distribution is approximately given by
\beq
\sigma(l)\simeq \sqrt{\frac{\mu /2}{ k(r+s)-c (k l_{peak} -1) }}
\eeq
Since the typical length is controlled by the fitness only, while the width is controlled also by the insertion rate, there are different regimes where the width is much smaller, comparable or larger than the average length.

The exception to the analysis in this section is the case when the fitness is always negative, i.e. $f(\max(l_0,\hat{l}))<0$. In this case, $l_{peak}=\hat{l}=l_0$, i.e. the plasmids do not tend to grow. Instead, they get eliminated from the bacterial populations.


\subsubsection*{Mobility and plasmid length}

The distribution (\ref{eq:p}) describes the length of plasmids with a fixed value of $l_c$, $r_c$ and $c_c$. However, there are several different types of conjugative plasmids with different conjugation machineries. Each of this class corresponds to a different set of parameters $l_c,r_c,c_c$ and have its own length distribution. Experimental evidence for these different distributions can be found in \cite{smillie2010mobility}. 

Mobilizable plasmids have relaxases but no MPF machinery. They have to rely on other conjugative plasmids present in the same host bacterium or on the host bacterium itself to infect other bacteria. The cost of this type of plasmids $c_c$ is much lower compared to conjugative ones, but their mobility $r_c$ is also greatly reduced. 

\begin{figure}
\begin{center}
\includegraphics[width=\columnwidth]{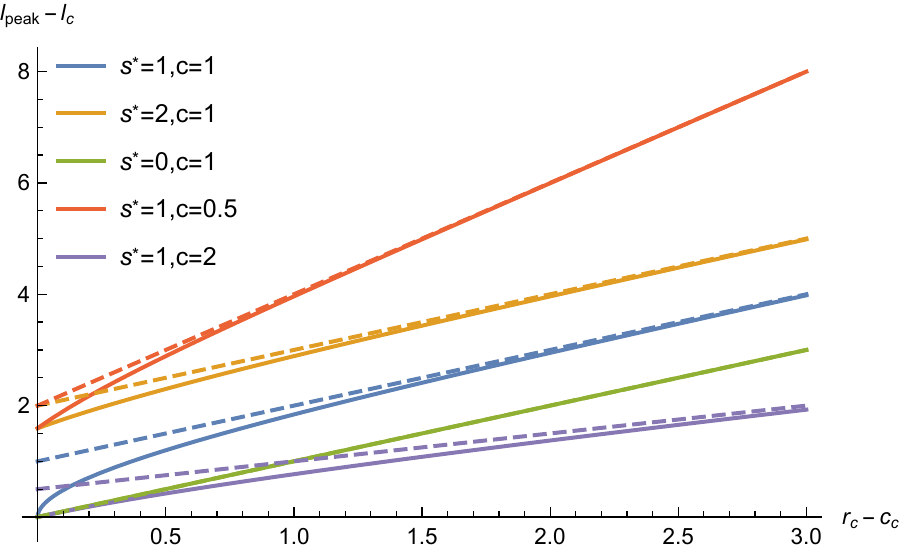}
\caption{Typical length of genes unrelated to mobility as a function of the difference between conjugation rate and cost $r_c-c_c$. The dashed lines represent the linear approximation.  The other parameters are $k=1$, $l_0=l_c$.}\label{fig_mob}
\end{center}
\end{figure}

We focus on the length of genes not related to mobility $l-l_c$. The fitness described by equation (\ref{eq_fitness0}) depends only on the difference between transmission rate and cost of the machinery, i.e. $r_c-c_c$. The typical length is
\beq
l_{peak}-l_c=\frac{s^*+r_c-c_c}{c}+\frac{1}{k}W\left(-\frac{ks^*}{c} e^{-\frac{k(s^*+r_c-c_c)}{c}}\right)
\eeq
which is an increasing function of $r_c-c_c$ and $s^*$ while decreasing in $k$, $c$. For $r_c-c_c$ not too small, its behaviour is approximately linear, as shown in Figure \ref{fig_mob}: $l_{peak}\simeq l_c+{(s^*+r_c-c_c)}/{c}$.

If conjugative plasmids were not born with the conjugation machinery, but obtained it through insertion, then their distribution would be given by equation (\ref{eq_fitness_p0}) where $p_0(l)$ would be the distribution of mobilizable plasmids. The typical length would be even longer in this case, since the initial length would be about the typical length of mobilizable plasmids: $l_{peak}\simeq l_{peak}^{mob}+l_c+{(s^*+r_c-c_c)}/{c}$. However, its dependence on $r_c-c_c$ would not change.

A clear prediction of our model is that the length of the plasmid (neglecting the genes of the machinery) increases with the efficiency of the transmission machinery. In particular, conjugative plasmids should be much longer than mobilizable plasmids.

\subsubsection*{The effect of increasing antibiotic concentration in the environment}

Antibiotics are one of the greatest achievements of modern medicine. Still their broad and sometimes unnecessary use represents a considerable hazard for public health as it may frustrate antibiotic effectiveness due to the increasing appearance of resistant bacteria.
This increase is apparent from the spreading of resistance genes. The increased antibiotic stress in the environment has important evolutionary effects \cite{andersson2014microbiological} also on plasmid length. 


The average fitness effect of a specific plasmid increases both with the frequency of encounters with antibiotic-rich environments and with the antibiotic concentrations. This is because on one side, higher concentrations increase the selective pressure on bacteria, while on the other side, selection occurs more often in time and space and therefore its average value over space and time increases.  This translates in an increase in $s^*$ in equation (\ref{eq_fitness0}), while the costs and the transmission do not change. The effect is an increase both in $s$ and in $r$.  

Antibiotic stress is a quite recent form of environmental stress that became quickly widespread not only for the broad medical prescription but especially for the extensive use of antibiotic at very low concentration in livestock breeding. 

\begin{figure}
\begin{center}
\includegraphics[width=\columnwidth]{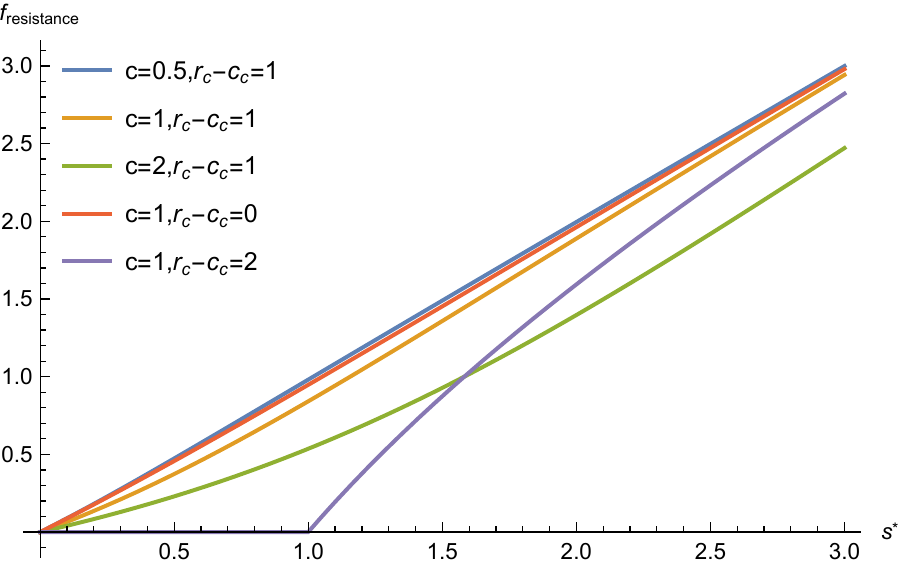}
\caption{Typical antibiotic resistance as a function of the environmental stress. 
The increase in antibiotic resistance is proportional to the increase in plasmid length. 
The other parameters are $k=1$, $l_c=0.3$, $l_0=l_c$.}\label{fig_anti}
\end{center}
\end{figure}

In Figure \ref{fig_anti} we show the effect of increasing environmental stress (such as growing antibiotic concentrations) on the typical length for several combinations of parameters. The current increase in selection strength is expected to result in a long-term increase in the typical plasmid length. This corresponds also to a proportional long-term proliferation of antibiotic resistance $\Delta f_{resistance}=c\Delta l_{peak}$.

\begin{figure}
\begin{center}
\includegraphics[width=\columnwidth]{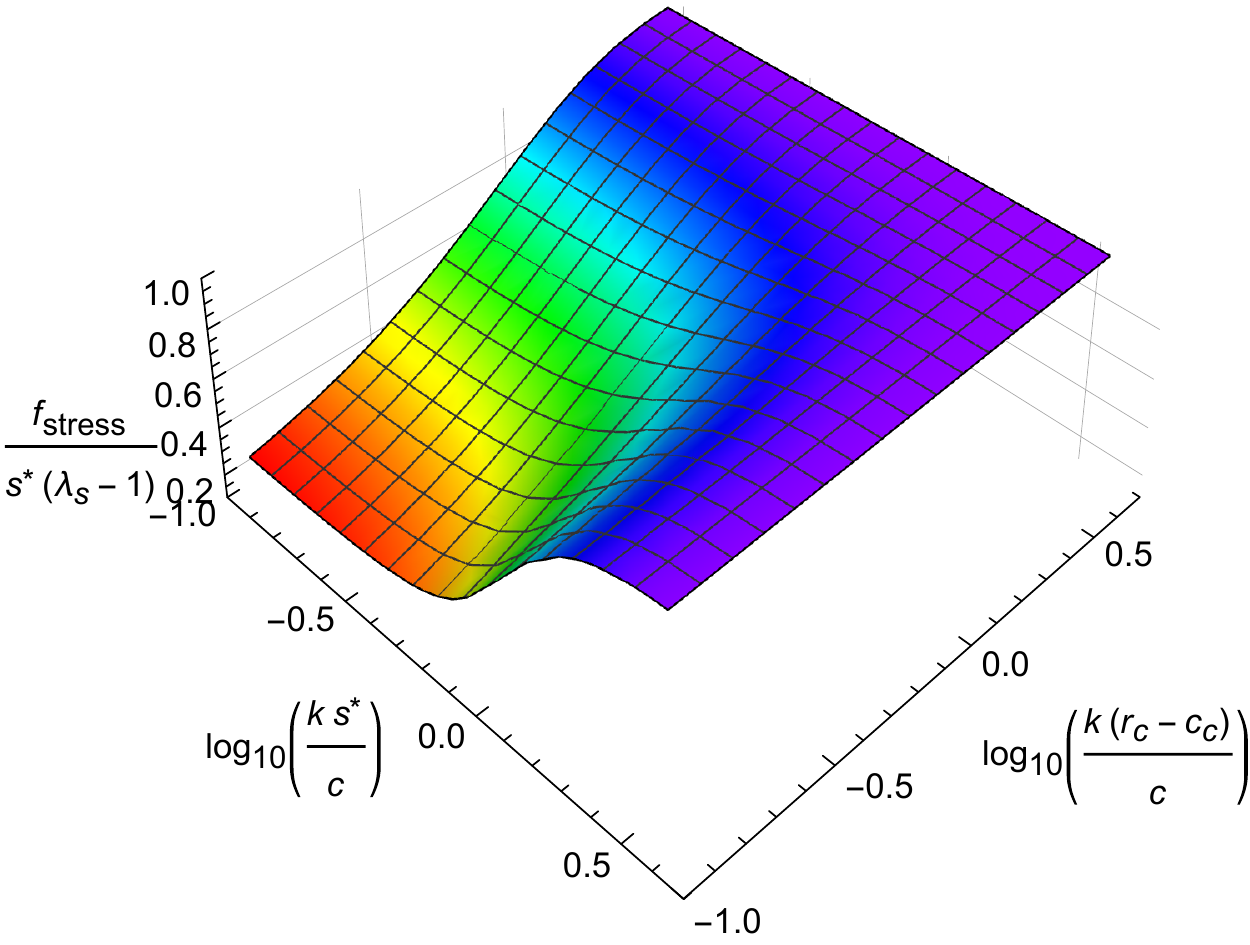}
\caption{Typical fitness of plasmids after a sudden increase in environmental stress, as a function of the selection/cost rate $ks^*/c$ and the mobility/cost rate $k(r_c-c_c)/c$, with $l_0=l_c$.}\label{fig_stress}
\end{center}
\end{figure}

However, rapid changes in environmental stress lead to a transient regime where the distribution is out of equilibrium. In particular, when the environmental stress induces an increase of a factor $\lambda_s$ in the selection coefficients, plasmids experience a typical increase in fitness of order
\beq
f_{stress}=(\lambda_s-1)\left[s^*+\frac{c}{k}W\left(-\frac{ks^*}{c} e^{-\frac{k(s^*+r_c-c_c)}{c}}\right)\right]
\eeq
(or $f_{stress}=(\lambda_s-1) s^*(1-e^{-kl_0})$ if $l_{peak}=l_0$). This transient, stress-induced fitness is illustrated in figure \ref{fig_stress} and it is ultimately responsible for the fast spreading of plasmid-mediated antibiotic resistance in antibiotics presence. 

\section*{Discussion}

In this paper we have proposed a model for plasmid fitness and evolution. 

Our results show that there is a significant load on plasmids due to 
redundant genes. These genes perform functions that could be advantageous for the bacteria. From the bacteria point of view each of these genes represents a cost, therefore for the host  bacteria a plasmid redundantly encoding for a function is less advantageous than a plasmid non-redundant for the same function. 
For this reason  plasmids with more redundancies will be disfavoured, while mechanism reducing redundancies will be favoured, as   
realised for example by the different types of Insertion Sequences that are exclusive among them \cite{carattoli2009resistance}.

Plasmids are  actively maintained by environmental stresses on bacterial populations. In absence of stress, these plasmids would be lost at a rate equal to their load, which would ultimately result in a complete loss of plasmids \cite{bichsel2010early, stewart1977population, hoeven1984mathematical,tazzyman2013fixation,levin1977probability}.

Interestingly, the model predicts that plasmids tend to be pushed by ``insertion pressure'' at the border of stability within host populations. This agrees with the fact that plasmids are widespread across bacterial and archaea populations, but there is no clear evidence of mobile plasmids being fixed, i.e. present in all individuals of the population. Near the border of stability, small perturbations in the fitness caused by changes in the environment may result in plasmid loss or establishment. It would be therefore difficult with this kind of perturbations to get to permanent plasmid establishment in the population.

We predict a positive correlation between mobility and length of plasmids. The number of genes not related with conjugation grows with the difference between the transmission rate and the cost of the plasmid. For the same reason, conjugative plasmids are predicted to contain more genes not related to the reproductive machinery than mobilizable plasmids. Our results agree with current evidence for the relation between plasmid conjugation rates and length \cite{smillie2010mobility}. 

Our results show also that the increase in antibiotic stress in the environment, for example the increase in antibiotic concentrations as a consequence of human use, increases both the length and the amount of antibiotic resistance genes encoded by plasmids. In presence of growing antibiotic concentrations, bacterial populations experience a sudden increase in selection for antibiotic resistance, which results in a positive contribution to plasmid fitness. During a transient period, this fitness contribution is not counterbalanced by the insertion load, therefore the plasmid population grows faster. This effect contributes to the spreading of antibiotic resistance in the bacterial population and ultimately to save the population from extinction \cite{bichsel2010early, stewart1977population, hoeven1984mathematical,tazzyman2013fixation,levin1977probability}. 

The limits of the model lie in some simplifications that we make. 
First, we assume that there are beneficial genes for several functions, but we neglect the differences in the benefits (and costs) of these genes. We do not consider the fitness and the full distribution of lengths for genes with different functions, or for opportunistic genes. 
Second, we do not take into account the impact of evolution on the average plasmid fitness. The average fitness of an evolved plasmid of length $l$ is larger than the average fitness of a random set of genes of length $l$, since evolution favour plasmids with less redundant genes. 
Third, our evolutionary model is very simple and does not include the possibility of deleting genes, which would result in a  reaction-drift-diffusion process. Moreover we neglect the complexity of the host bacterial population dynamics, saturation of plasmid population, incompatibility groups, evolution of the gene sequences, and stochastics effects like genetics drift. 

Future development should include features like gene functions and improved evolutionary and population genetic models, and inference of the parameters from available data. However, in its simplicity, this model captures some previously neglected aspects of plasmid evolution and allows to better understand the diffusion of plasmids and antibiotic resistance.\\

\acknowledgments

We thank Catherine Branger and Typhaine Billard-Pomares for useful discussions
. LF is funded by the grant ANR-12-JSV7-0007 from Agence Nationale de la Recherche, France. \\
\bibliographystyle{plain}
\bibliography{plasmidbib}

\appendix

\section{Exact solution for the stochastic model}

The evolution equation (\ref{eq:diffeq}) is a continuous approximation of the exact equations for the mean number of plasmids of length $l$, $n_l(t)$. Rescaling lengths such that $l_{gene}=1$, the length $l$ takes integer values; we also rescale times such that the rate of new plasmids is 1. The exact equations form a linear set:
\begin{align} 
&\frac{dn_{l_0}}{dt}=1-\mu_{l_0}n_{l_0}+f_{l_0}n_{l_0}\\
&\frac{dn_{l}}{dt}=\mu_{l-1}n_{l-1}-\mu_{l}n_{l}+f_{l}n_{l}\quad,\quad {l>l_0}
\end{align}
These equations can be solved exactly level by level as $n_{l_0}(t)=(e^{(f_{l_0}-\mu_{l_0})t}-1)/(f_{l_0}-\mu_{l_0})$, $n_l(t)=e^{(f_l-\mu_l)t}\int_0^td\tau e^{-(f_l-\mu_l)\tau} \mu_{l-1}n_{l-1}(t)$. 

If $\forall l\geq l_0$, we have $f_l<\mu_l$, then the system admits a stationary solution
\beq
n_l\propto \frac{1}{\mu_l}\prod_{l=l_0}^l \left(1-\frac{f_{l'}}{\mu_{l'}}\right)^{-1}
\eeq
and in the limit of small $f_l/\mu_l$ this solution converges to equation (\ref{eq:nl}). 

Otherwise, assuming that $f_l-\mu_l$ has a single maximum, we denote the length with the highest value of $f_l-\mu_l$ by $\hat{l}$, i.e. $f_{\hat{l}}-\mu_{\hat{l}}=\max_l(f_l-\mu_l)$. By considering only the leading terms, it is possible to integrate the above equations explicitly to get
\beq
n_l(t)\sim \exp[m_l t]\cdot \prod_{l'\leq l} \left| f_{l'}-\mu_{l'}-m_{l'-1}\right|^{-1} \\
\eeq
where $m_l=\max\left[0,\max_{\{l'\leq l\}}(f_{l'}-\mu_{l'})\right]$. 
Therefore the solution for $p_l=n_l/\sum_{l'}n_{l'}$ is
\beq
p_l\propto \begin{cases}0 & \mathrm{for}\ l<\hat{l}, \\  \prod_{l'=\hat{l}+1}^l (f_{\hat{l}}-\mu_{\hat{l}}-f_{l'}+\mu_{l})^{-1}& \mathrm{for}\ l\geq\hat{l}. \end{cases}
\eeq

\end{document}